\documentclass[11pt]{article}
\usepackage{amsmath,amssymb,amsthm,amsxtra,overpic,bbm,bm,epsfig,ulem,color,multirow}
\usepackage{braket}

\begin{document}

\begin{center}
{\Large Leptogenesis in the littlest inverse seesaw model}
\end{center}

\vspace{0.05cm}

\begin{center}
{\bf Yan Shao, \bf Zhen-hua Zhao\footnote{Corresponding author: zhaozhenhua@lnnu.edu.cn}, \bf Yi-Fei Duan,} \\
{ $^1$ Department of Physics, Liaoning Normal University, Dalian 116029, China \\
$^2$ Center for Theoretical and Experimental High Energy Physics, \\ Liaoning Normal University, Dalian 116029, China }
\end{center}

\vspace{0.2cm}

\begin{abstract}
The littlest inverse seesaw (LIS) model represents the first low-scale seesaw framework to successfully account for all six physical observables of the neutrino sector with merely two effective free parameters, making it highly worthy of in-depth investigation. In this work, we investigate realizations of leptogenesis in this framework. We consider two distinct scenarios. In the first, the two pseudo-Dirac sterile neutrino pairs are initially exactly degenerate and subsequently acquire small mass splittings via the RGE effects, enabling resonant leptogenesis to occur across different PD pairs and consequently enhancing leptogenesis. In the second, the two PD pairs feature a hierarchical mass spectrum, and leptogenesis proceeds via sterile neutrino oscillations through the ARS mechanism.
We show that the observed baryon asymmetry can be successfully reproduced in sizable regions of the parameter space without introducing additional free parameters, demonstrating that the LIS framework provides a viable and predictive setting for low-scale leptogenesis.
\end{abstract}

\newpage

\section{Introduction}

The existence of non-zero neutrino masses and the pronounced dominance of matter over antimatter in the Universe are among the most compelling indications of physics beyond the Standard Model (SM). On the one hand, neutrino oscillations have conclusively demonstrated that neutrinos have non-zero masses \cite{xing}, which contradicts the SM's assumption of massless neutrinos. On the other hand, the baryon asymmetry of Universe (BAU), quantified as \cite{planck}
\begin{eqnarray}
Y^{}_{\rm B} \equiv \frac{n^{}_{\rm B}-n^{}_{\rm \bar B}}{s} \simeq (8.69 \pm 0.04) \times 10^{-11}  \;,
\label{1.1}
\end{eqnarray}
where $n^{}_{\rm B}$ ($n^{}_{\rm \bar B}$) denotes the baryon (antibaryon) number density and $s$ the entropy density, remains unresolved.

A widely studied new physics model capable of addressing both problems is the type-I seesaw model \cite{seesaw1}-\cite{seesaw5}, which extends the SM by introducing right-handed neutrinos (RHNs) $N^{}_I$ (for $I=1, 2, ...$) with heavy Majorana masses. With the help of the seesaw mechanism, these states naturally give rise to light neutrino masses through their Yukawa interactions with the lepton doublets. In addition, their CP-violating and out-of-equilibrium decays can dynamically generate a lepton asymmetry in the early Universe, which is subsequently reprocessed into a baryon asymmetry via the sphaleron processes \cite{leptogenesis}-\cite{Lreview4}. However, this appealing simplicity entails a significant drawback: an extremely high mass scale. In the type-I seesaw model, reproducing sub-eV neutrino masses typically requires RHN masses close to ${\cal O}(10^{13})$ GeV. Furthermore, in scenarios with hierarchical RHNs, the Davidson-Ibarra bound implies that successful thermal leptogenesis requires the lightest RHN to be heavier than ${\cal O}(10^{9})$ GeV \cite{DI}. These exceedingly high scales are inaccessible to foreseeable experiments, driving the searches for alternative realizations of the seesaw framework.

In order to make the seesaw scale potentially accessible at present or future experiments, the inverse seesaw (ISS) model offers an attractive alternative \cite{ISS1, ISS2}. In addition to the RHNs, the model also introduces some other singlet fermions $S^{}_{I}$ (for $I=1, 2, ...$). And there are the following Lagrangian terms relevant for the generation of neutrino masses:
\begin{eqnarray}
-\mathcal{L}^{}_{\rm mass} \supset (Y^{}_{\nu})^{}_{\alpha I} \, \overline{L^{}_\alpha} \, \widetilde{H} \, N^{}_I + (M^{}_{\rm R})^{}_{IJ} \, \overline{N^c_{I}} \, S^c_{J} + \frac{1}{2} (\mu^{}_{\rm s})^{}_{IJ} \overline{S^{}_{I}} S^{c}_{J} + {\rm h.c.}
\label{1.2}
\end{eqnarray}
where $L^{}_\alpha$ (for $\alpha = e, \mu, \tau$) and $H$ (for $\widetilde{H} = {\rm i \sigma^{}_2} H^*$ with $\sigma^{}_2$ being the second Pauli matrix) respectively denote the lepton and Higgs doublets, and the superscript $c$ denotes the charge conjugated fields. And $Y^{}_\nu$, $M^{}_{\rm R}$ and $\mu^{}_{\rm s}$ are the Yukawa and mass matrices in the flavor space. After the Higgs field acquires its vacuum expectation value $v=174$ GeV, the Yukawa couplings will lead to some mass terms connecting the left-handed neutrinos $\nu^{}_\alpha$ and RHNs: $(M^{}_{\rm D})^{}_{\alpha I} = (Y^{}_{\nu})^{}_{\alpha I} v$. In the $(\nu_{\alpha}^c, N^{}_{I}, S_{I}^c)$ basis, the complete neutrino mass matrix is thus given by
\begin{eqnarray}
M^{}_{\nu NS} = \begin{pmatrix}
0 & M^{}_{\rm D} & 0 \\
M^{T}_{\rm D} & 0 & M^{}_{\rm R} \\
0 & M^{T}_{\rm R} & \mu^{}_{\rm s} \\
\end{pmatrix} \;.
\label{1.3}
\end{eqnarray}
In this model the lepton number is violated by the Majorana mass terms of $S^{}_I$, so the $\mu^{}_{\rm s}$ parameters can be naturally small in the sense of $'$t Hooft \cite{Hooft} (i.e., the lepton number symmetry would get restored in the limit of $\mu^{}_{\rm s} \rightarrow 0$). Under the natural condition of $\mu^{}_{\rm s} \ll M^{}_{\rm D} \ll M^{}_{\rm R}$, the light neutrino masses can be derived by performing a block diagonalization of Eq.~(\ref{1.3}):
\begin{equation}
M^{}_{\nu} \simeq M^{}_{\rm D} (M^{T}_{\rm R})^{-1}\mu^{}_{\rm s} M^{-1}_{\rm R} M^{T}_{\rm D} \;.
\label{1.4}
\end{equation}
We see that the smallness of neutrino masses can be naturally ascribed to small values of $\mu^{}_{\rm s}$, rendering superheavy RHNs no longer necessary. Therefore, the ISS model provides an attractive framework that can naturally explain the smallness of neutrino masses while accommodating relatively low-scale RHNs and relatively large neutrino Yukawa couplings simultaneously.

To achieve a more predictive low-energy neutrino phenomenology, Ref.~\cite{CarcamoHernandez:2019eme} proposed the Littlest Inverse Seesaw (LIS) model --- a minimal and predictive realization of the inverse seesaw mechanism that fuses the structural simplicity of the Littlest Seesaw (LS) model \cite{King:2013iva}-\cite{King:2018fqh} with the low-scale phenomenological advantages of ISS scenarios. In this model, which only introduces two RHNs and two $S^{}_I$ singlet fermions, the mass matrices in Eq.~(\ref{1.3}) take the following special patterns~\cite{CarcamoHernandez:2019eme}:
\begin{equation}
M^{}_{\rm D}= \left(
\begin{array}{cc}
0 & b \\
a & 3b \\
a & b
\end{array}
\right) \;,\hspace{0.7cm}
M^{}_{\rm R}= M^{}_{0}\left(
\begin{array}{cc}
1 & 0 \\
0 & z
\end{array}
\right) \;,\hspace{0.7cm}
\mu^{}_{\rm s} =\mu^{}_{0} \left(
\begin{array}{cc}
1 & 0 \\
0 & \omega
\end{array}
\right) \;,
\label{1.5}
\end{equation}
with $\omega =e^{{\rm i} \frac{2\pi}{3} }$ and the other parameters being free parameters. The resulting light active neutrino mass matrix derived from Eq.~(\ref{1.4}) is then given by
\begin{equation}
M^{}_{\nu} \simeq m^{}_{a}\left(
\begin{array}{ccc}
0 & 0 & 0 \\
0 & 1 & 1 \\
0 & 1 & 1
\end{array}
\right) +m^{}_{b}\omega \left(
\begin{array}{ccc}
1 & 3 & 1 \\
3 & 9 & 3 \\
1 & 3 & 1
\end{array}
\right)    \;,
\label{1.6}
\end{equation}
where $m^{}_a=\mu^{}_0 a^2_{} /M^2_0$ and $m^{}_b=\mu^{}_0 b^2/M^2_0 z^2 $ are two effective mass parameters. The numerical analysis indicates that, in the normal ordering (NO) case of light neutrino masses, taking $m^{}_{a}\simeq 26.57$ meV and $m^{}_{ b}\simeq 2.684$ meV can help us successfully reproduce the observed neutrino mass squared differences, neutrino mixing angles, and the Dirac CP phase  \cite{CarcamoHernandez:2019eme}.
This model therefore represents the first low-scale seesaw framework to successfully account for all six physical observables of the neutrino sector with merely two effective free parameters, making it highly worthy of in-depth investigation.

Motivated by the minimal setup and remarkable predictive power of the LIS model, in this work we investigate leptogenesis within this framework. The study is meaningful for the following reasons. First and foremost, leptogenesis in ISS scenarios is inherently a challenging problem (see the next paragraph for details) \cite{us}. Against this backdrop, the LIS model raises a critical and non-trivial question: can it successfully accommodate a leptogenesis scenario that reproduces the observed BAU?
If a viable leptogenesis mechanism can be indeed realized in the LIS model, this will not only establish its consistency with early-Universe physics but also further enhance its phenomenological predictivity by imposing additional constraints on its parameter space. It is therefore of crucial importance to examine the viability of leptogenesis in the LIS model and explore the parameter bounds that leptogenesis can impose on this framework.

In the ISS framework, the singlet fermions $N^{}_I$ and $S^{}_I$ (collectively termed sterile neutrinos) naturally form pseudo-Dirac (PD) pairs \cite{pseudo1}-\cite{pseudo4}. In the exact lepton-number-conserving limit with $\mu^{}_{\rm s}=0$, each pair constitutes an effective Dirac fermion, corresponding to two degenerate Majorana states of opposite CP phases; a small nonzero $\mu^{}_{\rm s}$ splits this state into two nearly degenerate Majorana eigenstates with a mass difference around $\mu^{}_{\rm s}$.
The quasi-degenerate sterile neutrinos in this setup naturally enable resonant enhancement of CP violation in leptogenesis \cite{resonant1, resonant2}, a feature expected to facilitate successful leptogenesis even for TeV-scale sterile neutrinos. However, detailed studies show that generic ISS models produce a baryon asymmetry at least a few orders of magnitude below the observed value \cite{us}, due to strong lepton asymmetry cancellation within each PD pair (see section~3) and severe washout from relatively large neutrino Yukawa couplings.
Several strategies to restore successful leptogenesis in the ISS framework have been proposed: Refs.~\cite{rescue1, rescue2} adopt the Akhmedov-Rubakov-Smirnov (ARS) mechanism \cite{ARS1, ARS2}, where lepton asymmetry arises from sterile neutrino oscillations; Refs.~\cite{rescue3}-\cite{rescue5} take advantage of that quasi-degeneracies between different PD pairs (so that resonant leptogenesis can occur across different PD pairs) can enhance leptogenesis in the ISS scenario.

Inspired by Refs.~\cite{rescue1, rescue2} and \cite{rescue3}-\cite{rescue5}, in this work we investigate the viability of leptogenesis within the LIS framework for the following two distinct scenarios. In the first, the two PD sterile neutrino pairs are initially exactly degenerate (corresponding to $z=1$ in Eq.~(\ref{1.5})) and subsequently acquire small mass splittings via renormalization group evolution (RGE) effects, enabling resonant leptogenesis to occur across different PD pairs and consequently enhancing leptogenesis. In the second, the two PD pairs feature a hierarchical mass spectrum (corresponding to $z\gg1$ in Eq.~(\ref{1.5})), and leptogenesis proceeds via sterile neutrino oscillations through the ARS mechanism.

The remainder of this paper is organized as follows. In the next section, we recapitulate some basic formulas of the ISS model and leptogenesis, which serve as the foundation for our subsequent study. In Sections~3 and 4, we investigate the viability of leptogenesis within the LIS framework for the two distinct scenarios outlined above. Finally, a summary of our key results is presented in Section 5.

\section{Some basic formulas of ISS model and leptogenesis}

In this section, we recapitulate some basic formulas of ISS model and leptogenesis as a basis of our study.

In order to facilitate the leptogenesis calculations, it is most convenient to work in the basis where the sterile-neutrino mass submatrix becomes diagonal, real, and positive. This can be achieved by applying a unitary rotation $V^{}_{\rm R}$ to the bottom-right block $ M^{}_{NS}$ of Eq.~(\ref{1.3}) \cite{Diag}, such that
\begin{eqnarray}
V^{T}_{\rm R} M^{}_{NS} V^{}_{\rm R} \approx
\begin{pmatrix}
V^{T}_{1}\left[-M^{}_{\rm R}+\frac{1}{2}  \mu^{}_{\rm s} \right] V^{}_{1} &  0  \\
0	&  V^{T}_{2}\left[M^{}_{\rm R}+\frac{1}{2}   \mu^{}_{\rm s} \right] V^{}_{2}
\end{pmatrix} \;,
\label{2.1}
\end{eqnarray}
where $V^{}_{1}$ and $V^{}_{2}$ are the unitary matrices for diagonalizing $-M^{}_{\rm R}+1/2  \mu^{}_{\rm s}$ and $M^{}_{\rm R}+1/2  \mu^{}_{\rm s}$, respectively. The approximate form of $V^{}_{\rm R}$ is given by
\begin{eqnarray}
V^{}_{\rm R}\approx\frac{1}{\sqrt{2}}
\begin{pmatrix}
\mathbf{1}+\frac{\mu^{}_{\rm s} {M^{-1}_{\rm R}}^{}}{4}	& \mathbf{1}-\frac{\mu^{}_{\rm s} {M^{-1}_{\rm R}}^{}}{4} \\
-\mathbf{1}+\frac{\mu^{}_{\rm s} {M^{-1}_{\rm R}}^{}}{4} & \mathbf{1}+\frac{\mu^{}_{\rm s} {M^{-1}_{\rm R}}^{}}{4}
\end{pmatrix}
\begin{pmatrix}
V^{}_{1} & 0 \\
0  & V^{}_{2}
\end{pmatrix}  \;.
\label{2.2}
\end{eqnarray}
Under this transformation, the full neutrino mass matrix acquires a block structure as
\begin{eqnarray}
M^{\prime}_{\nu NS} \simeq \begin{pmatrix}
0 & v Y^{\prime}_{\rm D} & v Y^{\prime}_{\rm L} \\
v \left(Y^{\prime}_{\rm D}\right)^{T} & V^{T}_{1}\left[-M^{}_{\rm R}+\frac{1}{2}  \mu^{}_{\rm s} \right] V^{}_{1} & 0 \\
v \left(Y^{\prime}_{\rm L}\right)^{T} & 0 & V^{T}_{2}\left[M^{}_{\rm R}+\frac{1}{2}   \mu^{}_{\rm s} \right] V^{}_{2} \end{pmatrix} \;,
\label{2.3}
\end{eqnarray}
where the rotated Yukawa couplings are expressed as \cite{Yv}
\begin{eqnarray}
Y^{\prime}_{\rm D} & \simeq & \frac{1}{\sqrt{2}\,v}\left[ M^{}_{\rm D} \left( \mathbf{1} + \frac{\mu^{}_{\rm s} {M^{-1}_{\rm R}}}{4} \right) V^{}_{1} \right] \;,
\nonumber\\
Y^{\prime}_{\rm L} & \simeq & \frac{1}{\sqrt{2}\,v}\left[ M^{}_{\rm D} \left( \mathbf{1} - \frac{\mu^{}_{\rm s} {M^{-1}_{\rm R}}}{4} \right)V^{}_{2} \right] \;.
\label{2.4}
\end{eqnarray}

In the low-scale (TeV or so) leptogenesis regime which we are interested in, all the three lepton flavors have become distinguishable from one another and the lepton asymmetries stored in them should be tracked separately \cite{flavor1, flavor2}. Moreover, since the sterile neutrinos appear in quasi-degenerate PD pairs, all of them contribute comparably to the final asymmetry. The resulting baryon asymmetry can thus be written as
\begin{eqnarray}
Y^{}_{\rm B}  = c Y^{}_{\rm L} = c r \sum^{}_\alpha \varepsilon^{}_{\alpha} \kappa^{}_\alpha \;,
\label{2.5}
\end{eqnarray}
where $c \simeq -1/3$ is the conversion efficiency from the lepton asymmetry to the baryon asymmetry via the sphaleron processes, $r \simeq 4 \times 10^{-3}$ measures the ratio of the equilibrium number density of sterile neutrinos to the entropy density, and $\varepsilon^{}_\alpha$ is a sum over the sterile neutrinos (i.e., $\varepsilon^{}_\alpha= \sum^{}_{I}\varepsilon^{}_{I \alpha}$) of the CP asymmetries for their decays:
\begin{eqnarray}
	\varepsilon^{}_{I \alpha } \equiv \frac{ \Gamma \left( N^{}_I \to L^{}_\alpha + H \right) - \Gamma \left( N^{}_I \to \overline{L}^{}_\alpha + \overline{H} \right) }
	{\sum^{}_{\alpha} \left[ \Gamma \left( N^{}_I \to L^{}_\alpha + H \right) + \Gamma \left( N^{}_I \to \overline{L}^{}_\alpha + \overline{H} \right) \right]  } \;.
\label{2.6}
\end{eqnarray}
Finally, $\kappa^{}_\alpha$ are the efficiency factors that take account of the washout effects due to the inverse decays of sterile neutrinos and various lepton-number-violating scattering processes. The explicit expressions for $\varepsilon^{}_{I \alpha}$ and $\kappa^{}_\alpha$ are given below.

In the resonant leptogenesis regime relevant for our work, which involves two pairs of PD sterile neutrinos (i.e., four nearly degenerate states), the CP asymmetries differ from the conventional two-state expressions \footnote{For the standard two-state resonant leptogenesis treatment, see e.g. Ref.~\cite{Jukkala:2021sku}. As demonstrated in Ref.~\cite{daSilva:2022mrx} (see Figure~1 therein), the two-state mixing approximation can lead to visibly different CP asymmetries compared to the full multi-state resummation in the presence of several nearly degenerate states. The CP asymmetries used in the present analysis is therefore more appropriate for the multi-state quasi-degenerate scenario studied here.}. In particular, the effective neutrino Yukawa couplings in our framework are given by the following form \cite{Pilaftsis:2003gt}-\cite{daSilva:2022mrx}
\begin{eqnarray}
\left( \bar{h}^{}_{+} \right)^{}_{\alpha I } = h_{\alpha I } + {\rm i} \mathcal{V}^{}_{\alpha I }
	- {\rm i} \sum_{J, K, L=1}^{4} h^{}_{\alpha J }  \mathcal{F}^{}_{IJKL}  \;,
\label{2.7}
\end{eqnarray}
where $h^{}_{}= (Y^{\prime}_{\rm D}, Y^{\prime}_{\rm L})$, and
\begin{eqnarray}
 	&&\hspace{-1.5cm} \mathcal{V}^{}_{\alpha I } =-\sum^{}_{\beta =e, \mu, \tau} \sum^{}_{J \neq I} \frac{h^{*}_{ \beta I } h^{}_{\beta J } h^{}_{\alpha J } }{16 \pi} f\left( \frac{M^2_{J}}{M^2_{I}} \right)  \;, \nonumber \\
	&&\hspace{-1.5cm} \mathcal{F}^{}_{IJKLMN} =\frac{1}{\left( M^{2}_{I} - M^{2}_{J} + 2 {\rm i} M^{2}_{I} A^{}_{JJ} \right) F^{(2)}_{IJKL}}  \times
\{ M^{}_{I}  \left( \mathcal{M}^{}_{IIJ } + \mathcal{M}^{}_{JJI} \right) \nonumber \\
	&&\hspace{0.6cm} -{\rm i} R^{}_{IL} \left[ \mathcal{M}^{}_{ILJ } \left( \mathcal{M}^{}_{IIL } + \mathcal{M}^{}_{LLI } \right) + \mathcal{M}^{}_{JJL} \left( \mathcal{M}^{}_{ILI } + \mathcal{M}^{}_{LIL} \right) \right]  \nonumber \\
	&&\hspace{0.6cm} -{\rm i} R^{}_{IK} \frac{1}{F^{(1)}_{IKL}} \left[ \mathcal{M}^{}_{IKJ } \left( \mathcal{M}^{}_{IIK } + \mathcal{M}^{}_{KKI } \right) + \mathcal{M}^{}_{JJK} \left( \mathcal{M}^{}_{IKI } + \mathcal{M}^{}_{KIK} \right) \right]
\} \;,
	\label{2.8}
\end{eqnarray}
with $f(x) = \sqrt{x} [ 1 - (1+x) \ln (1+1/x)]$ being the loop function, $\mathcal{M}^{}_{IJK} \equiv M^{}_{I} A^{}_{JK}$ and
\begin{eqnarray}
&& A^{}_{IJ} = \frac{ \left( h^{\dagger} h \right)^*_{IJ} }{16 \pi} \;, \hspace{1cm} R^{}_{IJ} = \frac{ M^{2}_{I} }{ M^{2}_{I} - M^{2}_{J} + 2 {\rm i} M^{2}_{I} A^{}_{JJ} }  \;,\nonumber \\
&& F^{(1)}_{IJK}= 1+ 2 {\rm i} {\rm Im} (R^{}_{IK}) T^{}_{IJK}\;,\nonumber \\
&& F^{(2)}_{IJKL}= F^{(1)}_{IJL}+ 2 {\rm i} {\rm Im} \left(R^{}_{IK} \frac{1}{F^{(1)}_{IKL}}\right) T^{}_{IJK}\;,\nonumber \\
&& T^{}_{IJK}= \frac{ M^{2}_{I} |A^{}_{JK}|^{2} + M^{}_{J}M^{}_{K}{\rm Re} A^{2}_{JK}}{ M^{2}_{I} - M^{2}_{J} + 2 {\rm i} M^{2}_{I} A^{}_{JJ} } \;.
\label{2.9}
\end{eqnarray}
The corresponding CP-conjugate effective Yukawa couplings $\left( \bar{h}^{}_{-} \right)^{}_{\alpha I }$ can be obtained by simply making the replacements $h^{}_{\alpha I } \to h^{*}_{\alpha I }$ in the above expressions. Furthermore, in terms of these effective Yukawa couplings, the decay widths of sterile neutrinos can be expressed as
\begin{eqnarray}
\Gamma \left( N^{}_I \to L^{}_{\alpha} + H \right) = \frac{ M^{}_{I} }{16\pi} \left| \left( \bar{h}^{}_{+} \right)^{}_{\alpha I } \right|^2 \;, \hspace{1cm}
\Gamma \left( N^{}_I \to \overline{L}^{}_\alpha + \overline{H} \right) = \frac{ M^{}_{I} }{16\pi} \left| \left( \bar{h}^{}_{-} \right)^{}_{\alpha I } \right|^2 \;.
\label{2.10}
\end{eqnarray}
Therefore, the flavor-specific CP asymmetries for the decays of the $I$-th sterile neutrino are given by
\begin{eqnarray}
	\varepsilon^{}_{I \alpha }
	= \frac{ \left| (\bar{h}^{}_{+})^{}_{\alpha I } \right|^2 - \left| (\bar{h}^{}_{-})^{}_{\alpha I } \right|^2  }
	{ \left( \bar{h}^{\dagger}_{+} \bar{h}^{}_{+} \right)^{}_{I I} + \left( \bar{h}^{\dagger}_{-} \bar{h}^{}_{-} \right)^{}_{II} } \;.
\label{2.11}
\end{eqnarray}

In conventional type-I seesaw frameworks, the interactions of unstable heavy neutrinos are effectively captured by $\Delta L = 2$ scattering processes of the form
$L_\alpha H \leftrightarrow L_\beta^c H^\dagger$, mediated by heavy neutrino exchanges. These scattering amplitudes can be decomposed into on-shell and off-shell contributions, with the on-shell part corresponding to inverse decay processes. The reaction density associated with the on-shell contribution can be written as
\begin{eqnarray}
\gamma_{\Delta L=2, I\alpha }^{\rm on} = \frac{\gamma^D_{ I\alpha }}{4}  \equiv
\frac{1}{4}\, n^{\rm eq}_{N^{}_I}\,\frac{\mathcal{K}_{1}}{\mathcal{K}_2}
\,\Gamma_{ I \alpha} =
\frac{4}{z} \frac{M^{4}_I}{246\pi^3}(|(\bar{h}^{}_{+})_{\alpha I }|^2+|(\bar{h}^{}_{-})_{\alpha I }|^2)
\mathcal{K}_{1}(z\, M^{}_I/M^{}_1) \;,
\label{2.12}
\end{eqnarray}
where $\mathcal{K}_{1,2}(z)$ are the modified Bessel functions, $z\equiv M_{1}/T$ with $T$ being the thermal bath temperature, and $\Gamma^{}_{ I\alpha }=\Gamma \left( N^{}_I \to L^{}_{\alpha} + H \right)+\Gamma \left( N^{}_I \to \overline{L}^{}_\alpha + \overline{H} \right)$. The strength of washout effects induced by inverse decays is commonly characterized by the washout parameters
\begin{eqnarray}
K^{}_{\alpha}= \sum^{}_{I} K_{ I \alpha} =  \sum^{}_{I} \frac{\Gamma^{}_{ I \alpha}}{H(T=M^{}_{I})} \;,
\label{2.13}
\end{eqnarray}
where the Hubble expansion rate is given by $H(T) \simeq 1.66 \sqrt{g_*}\, T^2/M^{}_{\rm Pl}$, with $g^{}_*$ denoting the effective number of relativistic degrees of freedom and $M^{}_{\rm Pl} \simeq 1.22\times 10^{19}$ GeV. A distinctive feature of the ISS framework is that lepton-number-violating washout effects are naturally suppressed in the limit of $\mu^{}_{\rm s} \to 0$. In this limit, lepton number is approximately conserved, and destructive interference between the two members of each PD pair leads to an almost complete cancellation of $\Delta L = 2$ washout contributions \cite{Yv, ISSSO10}.
As a consequence, the total washout rate is protected by an approximate symmetry and vanishes identically in the exact lepton-number-conserving limit, even if the naive washout parameters in Eq.~(\ref{2.13}) are large. Introducing a finite $\mu_{\rm s}$ explicitly breaks this symmetry, breaking the cancellation and reinstating effective inverse-decay washout. This demonstrates that interference effects, absent in standard seesaw models, play a central role in ISS leptogenesis. To properly account for these interference contributions, it is necessary to return to the fundamental definition of the full $\Delta L = 2$ scattering rate,
\begin{eqnarray}
 \gamma_{\Delta L=2,\alpha}^{\rm tot}=
\frac{1}{z}\frac{M_{1}^{4}}{64\,\pi^{4}}\,\int_{x_{\rm
thr}}^{\infty}dx\,\sqrt{x}\,\hat{\sigma}_{\Delta
L=2,\alpha}(x)\,\mathcal{K}_{1}(z\,\sqrt{x})\;,
\label{2.14}
\end{eqnarray}
where $x\equiv s/M^{2}_{N^{}_1}$, and $x^{}_{\rm  thr}$ denotes the kinematic threshold. The reduced $s$-channel cross section can be expressed as \cite{Pilaftsis:2005rv}
\begin{eqnarray}
\widehat{\sigma}^{}_{\Delta L=2,\alpha}(x)
\!\!&=&\!\!  \sum_{\beta=1}^{3}\ \sum_{I, J=1}^{4}\ {\rm Re}\, \Bigg\{\,
\bigg[\,(\bar{h}^{}_+)^*_{\alpha I}\,(\bar{h}_+)_{\alpha J}\,
(\bar{h}^{}_+)^*_{\beta I}\,(\bar{h}_+)_{\beta J}\: +\:
(\bar{h}^{}_-)^*_{\alpha I}\,(\bar{h}_-)_{\alpha J}\,
(\bar{h}^{}_-)^*_{\beta I}\,(\bar{h}_-)_{\beta J}\,\bigg]
\:\frac{x\,\sqrt{a_{I}a_{J} } }{4\pi P^{*}_{I}P_{J} }\ \Bigg\} \;,\nonumber \\
\label{2.15}
\end{eqnarray}
where $P_{I}^{-1}= \left(x-a^{}_{I}+i\sqrt{a^{}_{I} c^{}_{I}}\right)^{-1}$ is the Breit-Wigner $s$-channel
propagator, with $a_{I}=(M^{}_{I}/M^{}_{1})^{2}$ and $c_{I}=(\Gamma^{}_{I}/M^{}_{1})^{2}$.
A key merit of this formulation is that it automatically incorporates interference effects among different sterile neutrino states, enabling a consistent definition of the effective washout strength:
\begin{eqnarray}
K^{\rm eff}_{\alpha} \equiv K^{}_{\alpha} \cdot  \frac{\gamma_{\Delta L=2,\alpha}^{\rm tot}}{\gamma^D_{ \alpha}/4} \;,
\label{2.16}
\end{eqnarray}
where $\gamma^{D}_{\alpha}\equiv \sum_{I}^{} \gamma^{D}_{ I \alpha}/4$.

In most phenomenologically viable parameter regions, leptogenesis occurs in the strong washout regime ($K^{\rm eff}_{\alpha} \geqslant 1$). In this case, the efficiency factor follows the approximate relation
\begin{eqnarray}
\kappa^{}_{\alpha} \approx \frac{2}{z^{}_{B} K^{\rm eff}_{\alpha}}\left[1-e^{-\frac{1}{2}z^{}_{B} K^{\rm eff}_{\alpha}}\right] \;,
\label{2.16}
\end{eqnarray}
where $z_{B}$ is well approximated by \cite{zB}
\begin{eqnarray}
z_{B} \approx  1+\frac{1}{2}\ln\left[1+\frac{\pi {K^{\rm eff}_{\alpha}}^2}{1024}\left(\ln\frac{3125\pi {K^{\rm eff}_{\alpha}}^2}{1024}\right)^{5}\right] \;.
\label{2.17}
\end{eqnarray}
For completeness, we also take account of the weak washout regime ($K^{\rm eff}_{\alpha} < 1 $). Depending on whether $K^{}_{\alpha}$ is greater or less than 1, the efficiency factor behaves as
\begin{eqnarray}
\kappa^{}_{\alpha} \approx
\left\{
\begin{array}{ll}
\frac{1}{4} K^{\rm eff}_{\alpha} \left(\frac{3\pi}{2} - z^{}_{B}\right) \hspace{0.85cm} (K^{}_{\alpha} > 1) \;; \\
\frac{9\pi^{2}}{64} K^{\rm eff}_{\alpha} K^{}_{\alpha} \hspace{1.8cm} (K^{}_{\alpha} < 1) \;.
\end{array}
\right.
\label{2.18}
\end{eqnarray}

\section{Leptogenesis assisted by renormalization group evolution}

In this section, we study the scenario that the two PD sterile neutrino pairs in the LIS model are initially exactly degenerate (corresponding to $z=1$ in Eq.~(\ref{1.5})) and subsequently acquire small mass splittings via the RGE effects \cite{FJN}, enabling resonant leptogenesis to occur across different PD pairs and consequently enhancing leptogenesis \cite{rescue3}-\cite{rescue5}.

\subsection{Viability of leptogenesis}

First, with the help of Eq.~(\ref{2.11}), we show that the total CP asymmetry obtained after summing over sterile neutrino states is strongly suppressed in generic ISS models. To illustrate this point explicitly, let us take the first PD pair as example, consisting of the first and third sterile neutrinos. Because of the relation $V^{}_{1}={\rm i} V^{}_{2}$ (see Eq.~(\ref{2.1})), there exist the following correlations among the relevant couplings:
\begin{eqnarray}
&& h^{}_{\alpha 1 }=i h^{}_{\alpha 3 }\;,
\hspace{0.5cm} \mathcal{V}^{}_{\alpha 1 }=i \mathcal{V}^{}_{\alpha 3 }\;, \hspace{0.5cm}
(\mathcal{B}_{+})_{\alpha 1 }\simeq i (\mathcal{B}_{-})^{*}_{ \alpha 3}\;,\hspace{0.5cm}
(\mathcal{B}_{-})_{\alpha 1 }\simeq i (\mathcal{B}_{+})^{*}_{\alpha 3 }\;,
\label{2.19}
\end{eqnarray}
where $(\mathcal{B}_{+})_{\alpha I }=\sum_{J, K, L, M, N=1}^{4} h^{}_{\alpha J }  \mathcal{F}^{}_{IJKLMN}$  (while $(\mathcal{B}_{-})_{\alpha I }$ can be obtained by making the replacement $h^{}_{\alpha I } \to h^{*}_{\alpha I }$). These relations imply that the effective Yukawa couplings satisfy
\begin{eqnarray}
 \left| (\bar{h}^{}_{+})^{}_{\alpha 1 } \right|^2 \approx \left| (\bar{h}^{}_{-})^{}_{\alpha 3 } \right|^2 \;,
 \hspace{0.5cm}
 \left| (\bar{h}^{}_{-})^{}_{ \alpha 1} \right|^2 \approx \left| (\bar{h}^{}_{+})^{}_{\alpha 3 } \right|^2 \;,
\label{2.20}
\end{eqnarray}
which leads to an approximate cancellation between $\varepsilon^{}_{1\alpha}$ and $\varepsilon^{}_{3\alpha}$. As a result, the total CP asymmetry from each PD pair is strongly suppressed.

In order to enhance leptogenesis, we study the scenario that the two PD pairs are initially exactly degenerate and subsequently acquire small mass splittings via the RGE effects, enabling resonant leptogenesis to occur across different PD pairs. In the SM framework, the one-loop RGE equations of the neutrino Yukawa coupling matrix $Y^{}_\nu$ and the mass matrix $M^{}_{\rm R}$ are described by \cite{ISSRGE}
\begin{eqnarray}
&& 16\pi^2  \frac{{\rm d}Y^{}_\nu}{{\rm d}t}  =\left( \alpha^{}_\nu -\frac{3}{2} Y^{}_l Y^\dagger_l + \frac{3}{2} Y^{}_\nu Y^\dagger_\nu \right)Y^{}_\nu \; ,
 \nonumber \\
&& 16\pi^2  \frac{{\rm d}M^{}_{\rm R}}{{\rm d}t}  =  M^{}_{\rm R} Y^\dagger_\nu Y^{}_\nu  \; ,
\label{3.1}
\end{eqnarray}
where the running parameter is defined as $t = \ln \left(\mu/ \Lambda\right)$, with $\mu$ denoting the renormalization scale and $\Lambda$ the cut-off scale. The coefficient $\alpha^{}_\nu$ encodes gauge and Yukawa contributions and is given by
\begin{eqnarray}
\alpha^{}_\nu  =  {\rm tr} \left(3 Y^{}_u Y^\dagger_u + 3 Y^{}_d Y^\dagger_d + Y^{}_l Y^\dagger_l +  Y^{}_\nu Y^\dagger_\nu \right)-\frac{9}{20} g^2_1 - \frac{9}{4} g^2_2  \;,
\label{3.2}
\end{eqnarray}
where $g^{}_1$ and $g^{}_2$ are the $\rm U\left(1\right)^{}_{\rm Y}$ and $\rm SU\left(2\right)^{}_{\rm L}$ gauge coupling constants, and $Y^{}_{u,d,l}$ are respectively the Yukawa matrices for up-type quarks, down-type quarks and charged leptons.

To illustrate the impact of RGE-induced corrections prior to presenting our full numerical results, we focus on the diagonal entries of $M^{}_{\rm R}$. Integrating Eq.~(\ref{3.1}) for the $II$ component of $M^{}_{\rm R}$ yields an approximate relation between its value at the cut-off scale $\Lambda$ and that at the leptogenesis scale $M^{}_0$:
\begin{eqnarray}
	M^{}_{II} \left( M^{}_0 \right) \simeq M^{}_{II} \left( \Lambda \right)
	\left[1 - \frac{1}{16 \pi^2} \left(Y^{ \dagger}_{\nu} Y^{}_\nu \right)^{}_{II} \ln \left(  \frac{  \Lambda }{ M^{}_0 }  \right) \right] \;,
\label{3.3}
\end{eqnarray}
which implies that even initially degenerate sterile neutrino masses may acquire radiatively generated splittings at lower energies. In particular, the difference between two diagonal elements evolves as
\begin{eqnarray}
	M^{}_{II} \left( M^{}_0 \right) - M^{}_{JJ} \left( M^{}_0 \right)
	\simeq \frac{ M^{}_{I} \left( \Lambda \right) }{16 \pi^2}
	\left[ \left( Y^{ \dagger}_{\nu} Y^{}_\nu \right)^{}_{JJ} - \left( Y^{ \dagger}_{\nu} Y^{}_\nu \right)^{}_{II} \right]
	\ln \left( \frac{  \Lambda }{ M^{}_0 } \right) \;.
\label{3.4}
\end{eqnarray}
This expression demonstrates that the mass splitting between two PD pairs is primarily driven by the differences in their Yukawa couplings. Moreover, as the mass scale $\mu^{}_0$ decreases, the RGE induced mass splittings can become substantially amplified, driven by the accompanying increase in the neutrino Yukawa couplings $Y^{}_\nu$ as implied by Eq.~(\ref{1.4}). As a result, the RGE induced mass splittings can potentially exceed the scale of $\mu^{}_0$ itself. At this point, although the Yukawa coupling matrix still satisfies the relationship described in Eq.~(\ref{2.19}), the $(\mathcal{B}_{\pm})_{\alpha I }$ terms no longer do so. Consequently, the cancellation of CP asymmetries is weakened, allowing for the potential generation of a sufficiently large baryon asymmetry.
We further note that the RGE-induced off-diagonal elements of $M_{\rm R}$, which modify the mass splittings and mixing pattern among the sterile neutrinos, play a non-trivial role in this framework. These contributions are consistently included in our numerical computations.

\begin{figure*}
\centering
\includegraphics[width=5in]{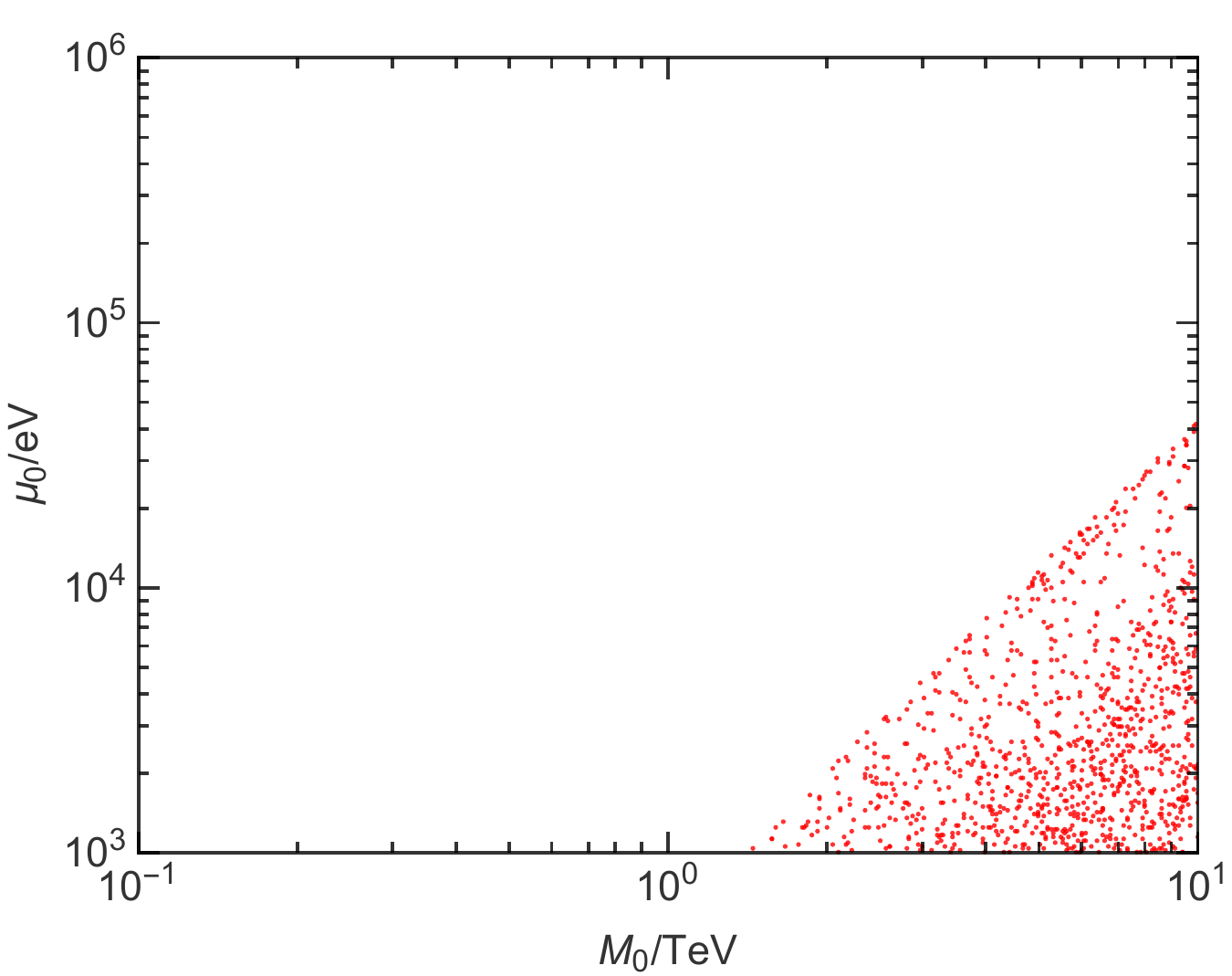}
\caption{ For the scenario studied in section~3, the parameter space of $M^{}_0$ versus $\mu^{}_0$ for successful leptogenesis. }
\label{fig1}
\end{figure*}

Our numerical analysis reveals that the observed value of $Y^{}_{\rm B}$ can be successfully reproduced. And Figure~\ref{fig1} has shown the parameter space of $M^{}_0$ versus $\mu^{}_0$ for successful leptogenesis. The points shown correspond to parameter sets for which the predicted $Y^{}_{\rm B}$ falls within its $3\sigma$ range, with the cut-off scale $\Lambda$ being scanned over the interval specified below.
These results are obtained for the following parameter settings. For the effective mass parameters $m^{}_a$ and $m^{}_b$ in Eq.~(\ref{1.6}), we take their values as mentioned below Eq.~(\ref{1.6}) which can help us successfully reproduce all six physical observables of the neutrino sector.
For $M^{}_0$, we allow it to vary in the range from the electroweak symmetry breaking (EWSB) scale up to $10$ TeV.
Then, inputting the values of $M^{}_0$, $m^{}_a$ and $m^{}_b$, the model parameters $a$ and $b$ in Eq.~(\ref{1.5}) can be obtained as functions of $\mu^{}_0$ with the help of their correlations below Eq.~(\ref{1.6}).
For the cut-off scale $\Lambda^{}_{}$, we allow them to vary in the range between $M^{}_0$ and the GUT scale $\sim 10^{16}_{}$ GeV.
To conform to the essential idea of the ISS model (i.e., $\mu^{}_{\rm s} \ll M^{}_{\rm D} \ll M^{}_{\rm R}$), in the calculations we have required $\mu^{}_{\rm s}$ to be at least two orders of magnitude smaller than $M^{}_{\rm D}$, and $M^{}_{\rm D}$ to be at least two orders of magnitude smaller than $M^{}_{\rm R}$.
The results demonstrate that in order to achieve a successful leptogenesis $M^{}_0$ should be within the range $1.5-10$ TeV and $\mu^{}_0$ should be within the range $1-40$ keV.
For smaller values of $M^{}_0$, relatively small values of $\mu^{}_0$ are needed to reproduce the observed value of $Y^{}_{\rm B}$.
As $M^{}_0$ increases, the allowed range of $\mu^{}_0$ that yields the observed $Y^{}_{\rm B}$ becomes broader.

%%%%%%%%%%%%%%%%%%%%%% FIG 1%%%%%%%%%%%%%%%%%%%%%%
%%%%%%%%%%%%%%%%%%%%%%%%%%%%%%%%%%%%%%%%%%%%%%%%%%

\subsection{Consequence for charged lepton flavor violation}

In this subsection, we investigate the charged lepton flavor violation (cLFV) processes in the LIS model under the constraints of successful leptogenesis.
Among various possible cLFV channels, the radiative decay $l^{}_\alpha \to l^{}_\beta \gamma $ is the most widely studied one, as it provides the most stringent experimental limits and is expected to exhibit the largest branching ratio. In particular, the MEG II experiment has established a strong upper bound on the decay ${\rm BR}(\mu \to e \gamma) < 1.5 \times 10^{-13}$ \cite{MEGII:2025gzr}.
Consequently, the evaluation of branching ratios for these decays within our setup serves as an essential consistency check on the viable parameter space. Under our setup, ${\rm BR}(l^{}_\alpha \to l^{}_\beta \gamma)$ can be calculated according to \cite{Forero:2011pc}
\begin{eqnarray}
{\rm BR}(l^{}_\alpha \to l^{}_\beta \gamma)= \frac{\alpha^3_W s^{2}_W}{256 \pi^2}\frac{m^{}_{l^{5}_\alpha}}{M^{4}_W}
\frac{1}{\Gamma^{}_{l^{}_{\alpha}}}|G^W_{\alpha \beta}|^2 \; ,
\label{3.5}
\end{eqnarray}
where
\begin{eqnarray}
&& G^{W}_{\alpha \beta}=\sum_{i=1}^7 K^*_{\alpha i} K^{}_{\beta i} G^W_\gamma \left(\frac{M^2_{i}}{M^{2}_W}\right) \; ,
 \nonumber \\
&& G_\gamma^W(x)=\frac{1}{12(1-x)^4}(10-43x+78x^2-49x^3+18x^3\ln{x}+4x^4) \;.
\label{3.6}
\end{eqnarray}
Here $ \alpha^{}_W \equiv g^2_W/(4\pi)$ denotes the weak fine-structure constant, $s^2_W= \sin^2 \theta^{}_W$ with $\theta^{}_W$ being the Weinberg angle, $M^{}_W$ the $W$-boson mass, $m^{}_{l^{}_\alpha}$ and $\Gamma^{}_{l^{}_{\alpha}}$ the mass and total decay width of the decaying charged lepton $l^{}_\alpha$.
Finally, one has the matrix $K=\left(K^{}_{L},K^{}_{H}\right)$ with
\begin{eqnarray}
&& K^{}_L = \left(I-\frac{1}{2}(M^{}_{\rm D}, 0)^*(M^{\prime -1}_{NS})^{*} M^{\prime -1}_{NS} (M^{}_{\rm D}, 0)^T\right) U^{\prime *} \; ,
 \nonumber \\
&& K^{}_H = (M^{}_{\rm D}, 0)^* (M^{\prime -1}_{NS})^{*} V \;,
\label{3.7}
\end{eqnarray}
where $M_{NS}^{\prime}$ is the sterile-neutrino mass matrix after taking account of the RGE effects.

\begin{figure*}
\centering
\includegraphics[width=5in]{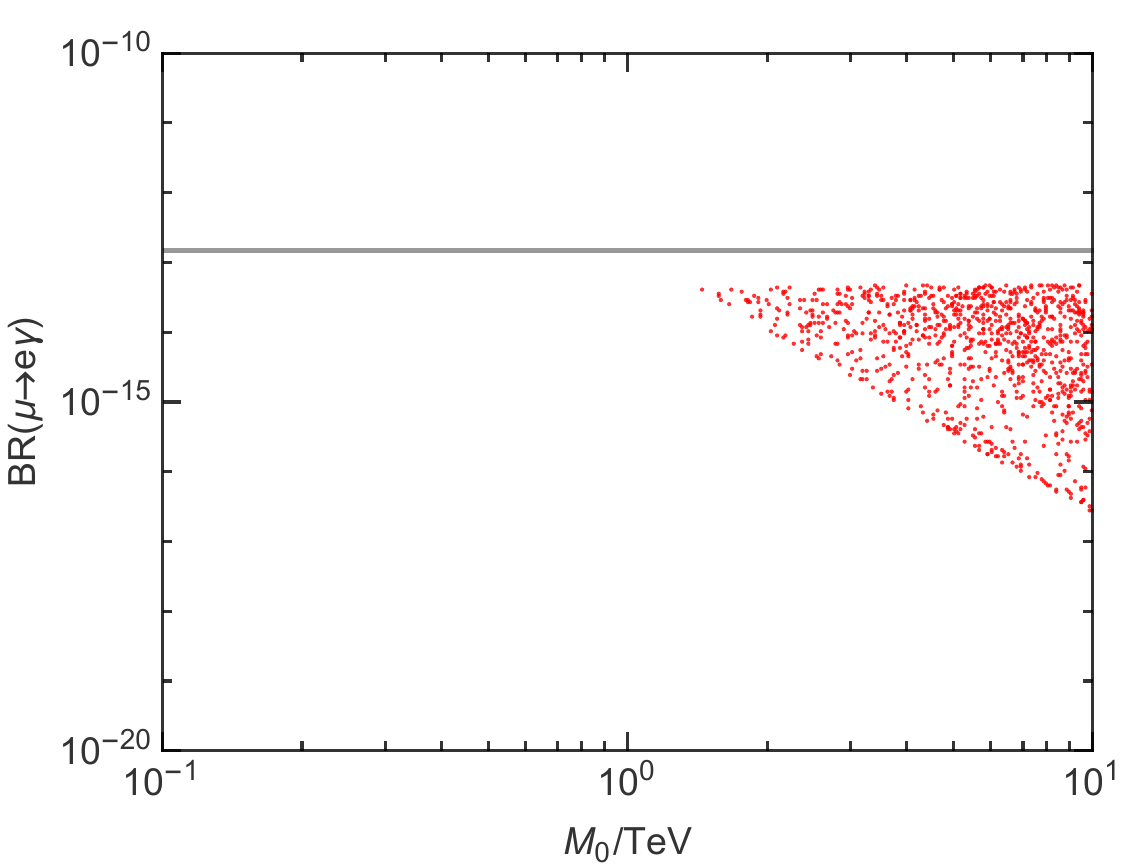}
\caption{ For the scenario studied in section~3.1, in the parameter space that allows for a reproduction of the observed value of $Y^{}_{\rm B}$, the allowed values of ${\rm BR}(\mu \to e \gamma)$ as functions of $M_0$. The horizontal line stands for the current upper bound on ${\rm BR}(\mu \to e \gamma)$. }
\label{fig2}
\end{figure*}

Now we investigate the consequences of the LIS model with respect to the $\mu \to e \gamma $ process under the constraints of successful leptogenesis. In Figure~\ref{fig2}, we have shown the allowed values of ${\rm BR}(\mu \to e \gamma)$ as functions of $M_0$. With the same parameter settings as in section~3.1, these results are obtained within the parameter space that allows for a successful reproduction of the observed baryon asymmetry of the Universe.
It turns out that, for the parameter space consistent with the observed value of $Y_{\rm B}$, the allowed values of ${\rm BR}(\mu \to e \gamma)$ can lie just below the current upper bound, placing this decay channel within the reach of upcoming high-sensitivity cLFV experiments such as MEG II, which is expected to improve the sensitivity to ${\rm BR}(\mu \to e \gamma)$ down to about $6 \times 10^{-14}$ \cite{MEGII:2018kmf}. Other planned experiments, such as Mu3e \cite{Mu3e:2020gyw} ($\mu \to 3e$) and Mu2e/COMET \cite{Mu2e:2014fns, COMET:2009qeh} ($\mu$--$e$ conversion), may also probe relevant regions of the parameter space through complementary cLFV channels.

%%%%%%%%%%%%%%%%%%%%%% FIG 2%%%%%%%%%%%%%%%%%%%%%%
%%%%%%%%%%%%%%%%%%%%%%%%%%%%%%%%%%%%%%%%%%%%%%%%%%

\section{Leptogenesis via ARS mechanism}

Noteworthy, the baryon asymmetry does not necessarily have to originate from the decay (freeze-out) processes of sterile neutrinos. For GeV scale sterile neutrinos, the baryon asymmetry instead originates from their production (freeze-in) processes. This scenario is referred to as leptogenesis via oscillations or ARS leptogenesis \cite{ARS1, ARS2}. Inspired by Refs.~\cite{rescue1, rescue2}, in this section we study the scenario that the two PD pairs feature a hierarchical mass spectrum (corresponding to $z \gg 1$ in Eq.~(\ref{1.5})), and leptogenesis proceeds via sterile neutrino oscillations through the ARS mechanism.
Under this setup, the baryon asymmetry is mainly generated by the oscillations of the lighter PD sterile neutrino pair, while the mass scale of the second PD pair is taken to be much heavier so that it effectively decouples during leptogenesis \cite{rescue2}.

In this scenario, the semi-classical Boltzmann equation approach to thermal leptogenesis is insufficient, thus it is necessary to solve a set of quantum kinetic equations for the density matrices of sterile neutrinos.
Due to the large dimensionality of the parameter space and complicated dynamics, it is difficult to provide a general analytical formula for the baryon asymmetry as a function of the model parameters. In the relevant calculations that follow, we make use of the latest version of the ULYSSES Python package \cite{uly} to obtain our results.

\begin{figure*}
\centering
\includegraphics[width=5in]{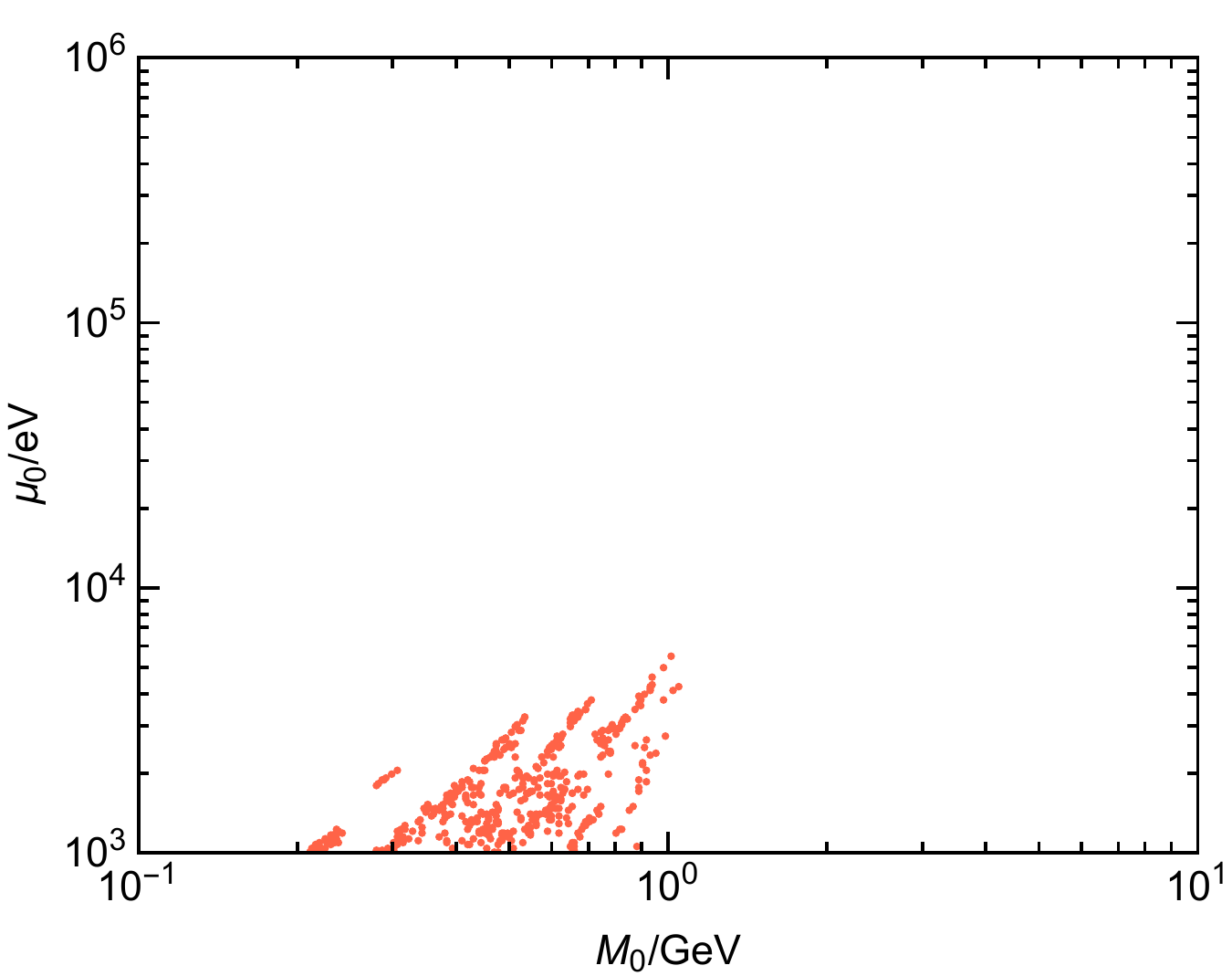}
\caption{ For the scenario studied in section~4, the parameter space of $M^{}_0$ versus $\mu^{}_0$ for successful leptogenesis. }
\label{fig3}
\end{figure*}

\begin{figure*}
\centering
\includegraphics[width=5in]{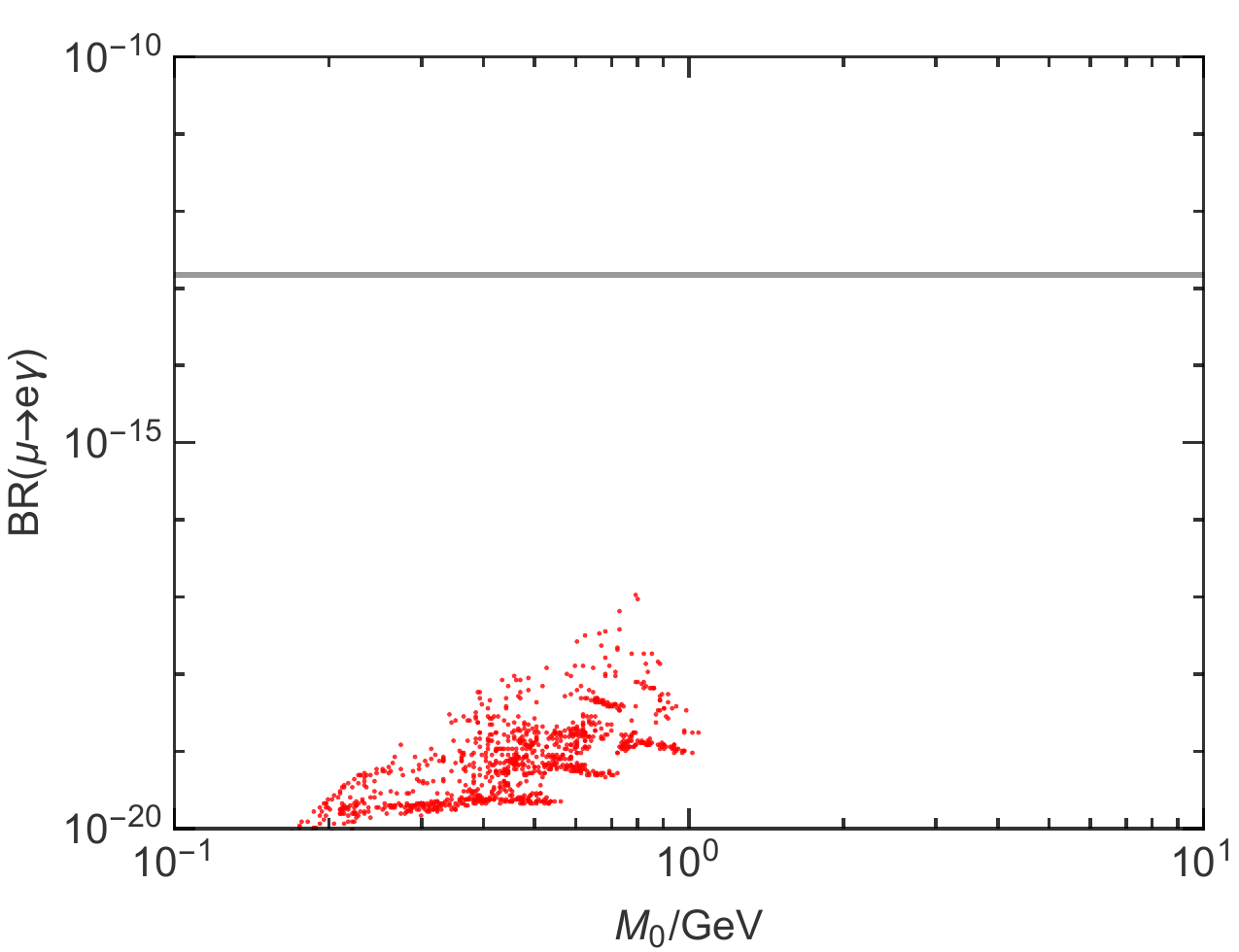}
\caption{ For the scenario studied in section~4, in the parameter space that allows for a reproduction of the observed value of $Y^{}_{\rm B}$, the allowed values of ${\rm BR}(\mu \to e \gamma)$ as functions of $M_0$. The horizontal line stands for the current upper bound on ${\rm BR}(\mu \to e \gamma)$. }
\label{fig4}
\end{figure*}

For this scenario, our numerical analysis reveals that the observed value of $Y^{}_{\rm B}$ can also be successfully reproduced.
And Figure~\ref{fig3} has shown the parameter space of $M^{}_0$ versus $\mu^{}_0$ for successful leptogenesis. The points displayed correspond to parameter sets for which the predicted $Y_{\rm B}$ falls within its $3\sigma$ range.
These results are obtained for the following parameter settings. For the parameter $z$, we adopt $z= 10$ as a benchmark value, ensuring that the heavier PD pair is sufficiently heavy to effectively decouple from leptogenesis. The mass scale $M^{}_0$ is varied between $0.1$ and $10$ GeV, placing the sterile neutrinos in the mass range relevant for the ARS leptogenesis scenario. The remaining parameter settings are the same as those in the last paragraph of Section 3.1.
The results show that the observed value of $Y^{}_{\rm B}$ can be successfully reproduced for $M^{}_0$ from $0.2$ to $1$ GeV and $\mu^{}_0$ from $1$ to $5$ keV.
Comparing this viable parameter region with the one obtained for the resonant leptogenesis scenario ($z=1$) in section~3.1, we see that the two allowed $M_0$--$\mu_0$ ranges appear completely disjoint. This is expected, since the resonant ($z=1$) and ARS ($z\gg1$) cases correspond to dynamically distinct regimes of the model, characterized by opposite limits of the parameter $z$, rather than being connected by a continuous variation of this parameter. In the resonant case, all four sterile states participate in the asymmetry generation through multi-state interference effects, while in the ARS case the heavier PD pair effectively decouples and the dynamics reduces to a two-state oscillation scenario.
For completeness, we have also examined the opposite limit $z \ll 1$. In this regime, the parameter $b$ in Eq.~(\ref{1.5}) is proportional to $z$ and therefore becomes significantly suppressed compared to the $z \gg 1$ case. In this regime, leptogenesis in the ARS mechanism is mainly driven by the oscillations of the lighter PD pair, whose relevant Yukawa couplings are proportional to $b$. As a result, the baryon asymmetry obtained in this regime is about two orders of magnitude below the observed value, and therefore cannot realize a successful leptogenesis.
For this scenario, we have also investigated the consequences of the model with respect to the $\mu \to e \gamma $ process. The corresponding results are shown in Figure~\ref{fig4}. It turns out that, for the parameter space consistent with the observed value of $Y^{}_{\rm B}$, the allowed values of ${\rm BR}(\mu \to e \gamma)$ are far below the current upper bound.
In contrast, the predicted branching ratios in the ARS scenario lie orders of magnitude below the current experimental limits and are expected to remain inaccessible to the foreseeable cLFV experiments.

%%%%%%%%%%%%%%%%%%%%%% FIG 3%%%%%%%%%%%%%%%%%%%%%%
%%%%%%%%%%%%%%%%%%%%%%%%%%%%%%%%%%%%%%%%%%%%%%%%%%

%%%%%%%%%%%%%%%%%%%%%% FIG 4%%%%%%%%%%%%%%%%%%%%%%
%%%%%%%%%%%%%%%%%%%%%%%%%%%%%%%%%%%%%%%%%%%%%%%%%%

\section{Summary}

The type-I seesaw model provides a well-motivated explanation for both the origin of neutrino masses and the baryon asymmetry of the Universe. However, in its conventional realization, the new particle states responsible for these phenomena typically reside at very high mass scales, placing them far beyond the reach of foreseeable experiments. The ISS model offers an appealing low-scale alternative, in which the smallness of neutrino masses arises naturally and sterile neutrinos may appear at experimentally accessible scales.

The LIS model represents the first low-scale seesaw framework to successfully account for all six physical observables of the neutrino sector with merely two effective free parameters, making it highly worthy of in-depth investigation. Motivated by the minimal setup and remarkable predictive power of this model, in this work we have investigated leptogenesis within this framework. The study is meaningful in the sense that successful leptogenesis in generic ISS models is highly challenging. If a viable leptogenesis mechanism can be indeed realized in the LIS model, this will not only establish its consistency with early-Universe physics but also further enhance its phenomenological predictivity by imposing additional constraints on its parameter space.

We have considered two distinct scenarios. In the first, the two PD sterile neutrino pairs are initially exactly degenerate (corresponding to $z=1$ in Eq.~(\ref{1.5})) and subsequently acquire small mass splittings via the RGE effects, enabling resonant leptogenesis to occur across different PD pairs and consequently enhancing leptogenesis. In the second, the two PD pairs feature a hierarchical mass spectrum (corresponding to $z\gg1$ in Eq.~(\ref{1.5})), and leptogenesis proceeds via sterile neutrino oscillations through the ARS mechanism. We find that the observed baryon asymmetry of the Universe can be successfully reproduced in both scenarios, demonstrating that the LIS model provides a viable setting for realizing low-scale leptogenesis.

In the quasi-degenerate case, successful leptogenesis is achieved through RGE-induced mass splittings.
We find that the observed baryon asymmetry can be successfully reproduced for $M^{}_0$ in the range $1.5-10$ TeV and $\mu^{}_0$ in the range $1-40$ keV.
For smaller values of $M^{}_0$, relatively small values of $\mu^{}_0$ are needed to reproduce the observed value of $Y^{}_{\rm B}$.
As $M^{}_0$ increases, the allowed range of $\mu^{}_0$ that yields the observed $Y^{}_{\rm B}$ becomes broader.
We have also examined the implications of the model for the lepton-flavor-violating process $\mu \to e \gamma$ within the parameter space consistent with successful leptogenesis, finding that the predicted branching ratios can be just below the current experimental upper bound by just one order of magnitude in certain parameter region.
This scenario may be tested by forthcoming high-sensitivity cLFV experiments such as MEG II, which are expected to probe the predicted branching ratios.

In the hierarchical scenario, leptogenesis proceeds via sterile neutrino oscillations through the ARS mechanism.
We find that the observed baryon asymmetry can be successfully reproduced for $M^{}_0$ from $0.2$ to $1$ GeV and $\mu^{}_0$ from $1$ to $5$ keV. In this scenario, within the parameter space consistent with successful leptogenesis, the predicted branching ratio ${\rm BR}(\mu \to e \gamma)$ are far below the current experimental upper bound.
In contrast, the corresponding cLFV signals in the ARS scenario are expected to remain experimentally inaccessible in the foreseeable future.

\vspace{0.5cm}

\underline{Acknowledgments} \vspace{0.2cm}

This work was supported in part by the National Natural Science Foundation of China under Grant No. 12475112, Liaoning Revitalization Talents Program under Grant No. XLYC2403152, and the Basic Research Business Fees for Universities in Liaoning Province under Grant No. LJ212410165050.


\begin{thebibliography}{99}

\bibitem{xing} Z. Z. Xing, Phys. Rep. {\bf 854}, 1 (2020).

\bibitem{planck} P. A. R. Ade {\it et al.} (Planck Collaboration), Astron. Astrophys. A {\bf16}, 571 (2014).

\bibitem{seesaw1} P. Minkowski, Phys. Lett. B {\bf 67}, 421 (1977).

\bibitem{seesaw2} M. Gell-Mann, P. Ramond and R. Slansky, in Supergravity, edited by P. van Nieuwenhuizen and D. Freedman, (North-Holland, 1979), p. 315.

\bibitem{seesaw3}  T. Yanagida, in Proceedings of the Workshop on the Unified Theory and the Baryon Number in the Universe, edited by O. Sawada and A. Sugamoto (KEK Report No. 79-18, Tsukuba, 1979), p. 95.

\bibitem{seesaw4} R. N. Mohapatra and G. Senjanovic, Phys. Rev. Lett. {\bf 44}, 912 (1980).

\bibitem{seesaw5} J. Schechter and J. W. F. Valle, Phys. Rev. D {\bf22}, 2227 (1980).

\bibitem{leptogenesis} M. Fukugita and T. Yanagida, Phys. Lett. B {\bf 174}, 45 (1986).

\bibitem{Lreview1} W. Buchmuller, R. D. Peccei and T. Yanagida, Ann. Rev. Nucl. Part. Sci. {\bf 55}, 311 (2005).

\bibitem{Lreview2} W. Buchmuller, P. Di Bari and M. Plumacher, Annals Phys. {\bf 315}, 305 (2005).

\bibitem{Lreview3} S. Davidson, E. Nardi and Y. Nir, Phys. Rept. {\bf 466}, 105 (2008).

\bibitem{Lreview4} D. Bodeker and W. Buchmuller, Rev. Mod. Phys. {\bf 93}, 035004 (2021).

\bibitem{DI} S. Davidson and A. Ibarra, Phys. Lett. B {\bf 535}, 25 (2002).

\bibitem{ISS1} D. Wyler and L. Wolfenstein, Nucl. Phys. B {\bf 218}, 205 (1983).

\bibitem{ISS2} R. N. Mohapatra, Phys. Rev. Lett. {\bf 56}, 561 (1986).

\bibitem{Hooft} G. $'$t Hooft, NATO Sci. Ser. B {\bf 59}, 135 (1980).

\bibitem{CarcamoHernandez:2019eme} A.~E.~C{\'a}rcamo Hern{\'a}ndez and S.~F.~King, Nucl. Phys. B \textbf{953}, 114950 (2020).

\bibitem{King:2013iva} S.~F.~King, JHEP \textbf{07}, 137 (2013).

\bibitem{Bjorkeroth:2014vha} F.~Bj{\"o}rkeroth and S.~F.~King, J. Phys. G \textbf{42}, 125002 (2015).

\bibitem{King:2015dvf} S.~F.~King, JHEP \textbf{02}, 085 (2016).

\bibitem{Bjorkeroth:2015ora} F.~Bj{\"o}rkeroth, F.~J.~de Anda, I.~de Medeiros Varzielas and S.~F.~King, JHEP \textbf{06}, 141 (2015).

\bibitem{Bjorkeroth:2015tsa} F.~Bj{\"o}rkeroth, F.~J.~de Anda, I.~de Medeiros Varzielas and S.~F.~King, JHEP \textbf{10}, 104 (2015).

\bibitem{King:2016yvg} S.~F.~King and C.~Luhn, JHEP \textbf{09}, 023 (2016).

\bibitem{Ballett:2016yod} P.~Ballett, S.~F.~King, S.~Pascoli, N.~W.~Prouse and T.~Wang, JHEP \textbf{03}, 110 (2017).

\bibitem{King:2018fqh} S.~F.~King, S.~Molina Sedgwick and S.~J.~Rowley, JHEP \textbf{10}, 184 (2018).

\bibitem{us} M.~J.~Dolan, T.~P.~Dutka and R.~R.~Volkas, JCAP {\bf 06}, 012 (2018).


\bibitem{pseudo1} L. Wolfenstein, Nucl. Phys. B {\bf 186}, 147 (1981).

\bibitem{pseudo2} S. T. Petcov, Phys. Lett. B 110, 245 (1982).

\bibitem{pseudo3} J. W. F. Valle and M. Singer, Phys. Rev. D 28, 540 (1983).

\bibitem{pseudo4} M. Kobayashi and C. S. Lim, Phys. Rev. D 64, 013003 (2001).

\bibitem{resonant1} A. Pilaftsis, Phys. Rev. D {\bf 56}, 5431 (1997).

\bibitem{resonant2} A. Pilaftsis and T. E. J. Underwood, Nucl. Phys. B {\bf 692}, 303 (2004).

\bibitem{rescue1} A. Abada, G. Arcadi, V. Domcke and M. Lucente, JCAP {\bf 11}, 041 (2015).

\bibitem{rescue2} A. Abada, G. Arcadi, V. Domcke and M. Lucente, JCAP {\bf 12}, 024 (2017).

\bibitem{ARS1} E. K. Akhmedov, V. A. Rubakov and A. Y. Smirnov, Phys. Rev. Lett. {\bf 81}, 1359 (1998).

\bibitem{ARS2} T. Asaka and M. Shaposhnikov, Phys. Lett. B {\bf 620}, 17 (2005).

\bibitem{rescue3}K.~Agashe, P.~Du, M.~Ekhterachian, C.~S.~Fong, S.~Hong and L.~Vecchi, JHEP \textbf{04}, 029 (2019).

\bibitem{rescue4} A. Mukherjee and A. K. Saha, Phys. Lett. B {\bf 849}, 138474 (2024).

\bibitem{rescue5} Y. Shao and Z. H. Zhao, Phys. Rev. D {\bf 113}, 015001 (2026).


\bibitem{Diag} B.~Karmakar and A.~Sil, Phys. Rev. D {\bf 96}, 015007 (2017).

\bibitem{Yv} S.~Blanchet, T.~Hambye and F.~X.~Josse-Michaux, JHEP {\bf 04}, 023 (2010).

\bibitem{flavor1} A. Abada, S. Davidson, F. X. Josse-Michaux, M. Losada and A. Riotto, JCAP {\bf0604}, 004 (2006).

\bibitem{flavor2} E. Nardi, Y. Nir, E. Roulet and J. Racker, JHEP {\bf0601}, 164 (2006).

\bibitem{Jukkala:2021sku} H.~Jukkala, K.~Kainulainen and P.~M.~Rahkila, JHEP \textbf{09}, 119 (2021).

\bibitem{Pilaftsis:2003gt} A.~Pilaftsis and T.~E.~J.~Underwood, Nucl. Phys. B \textbf{692}, 303-345 (2004).

\bibitem{Pilaftsis:2005rv} A.~Pilaftsis and T.~E.~J.~Underwood, Phys. Rev. D \textbf{72}, 113001 (2005).

\bibitem{daSilva:2022mrx} P.~C.~da Silva, D.~Karamitros, T.~McKelvey and A.~Pilaftsis, JHEP \textbf{11}, 065 (2022).

\bibitem{ISSSO10}S. Blanchet, P. S. Bhupal Dev and R. N. Mohapatra, Phys. Rev. D {\bf 82}, 115025 (2010).

\bibitem{zB} W.~Buchmuller, P.~Di Bari and M.~Plumacher, Annals Phys. {\bf 315}, 305 (2005).

\bibitem{FJN} R. Gonzalez Felipe, F. R. Joaquim and B. M. Nobre, Phys. Rev. D {\bf 70}, 085009 (2004).

\bibitem{ISSRGE} J.~Bergstrom, M.~Malinsky, T.~Ohlsson and H.~Zhang, Phys. Rev. D {\bf 81}, 116006 (2010).

\bibitem{MEGII:2025gzr} K.~Afanaciev \textit{et al.} [MEG II], Eur. Phys. J. C \textbf{85}, 1177 (2025).

\bibitem{Forero:2011pc} D.~V.~Forero, S.~Morisi, M.~Tortola and J.~W.~F.~Valle, JHEP \textbf{09}, 142 (2011).

\bibitem{MEGII:2018kmf} A.~M.~Baldini \textit{et al.} [MEG II], Eur. Phys. J. C \textbf{78}, 380 (2018).

\bibitem{Mu3e:2020gyw} K.~Arndt \textit{et al.} [Mu3e], Nucl. Instrum. Meth. A \textbf{1014}, 165679 (2021).

\bibitem{Mu2e:2014fns} L.~Bartoszek \textit{et al.} [Mu2e], [arXiv:1501.05241 [physics.ins-det]].

\bibitem{COMET:2009qeh} Y.~G.~Cui \textit{et al.} [COMET], KEK-2009-10.

\bibitem{uly} A. Granelli, C. Leslie, Y. F. Perez-Gonzalez, H. Schulz, B. Shuve, J. Turner and R. Walker, Comput. Phys. Commun. {\bf 291}, 108834 (2023).






\end{thebibliography}
\end{document}